\documentclass[conference]{IEEEtran}
\IEEEoverridecommandlockouts
\usepackage{cite}
\usepackage{amsmath,amssymb,amsfonts}
\usepackage{algorithmic}
\usepackage{graphicx}
\usepackage{textcomp}
\usepackage{xcolor}
\usepackage{subfigure}
\usepackage{bm}
\usepackage{floatrow}
\usepackage{stfloats}
\usepackage{url}
\def\BibTeX{{\rm B\kern-.05em{\sc i\kern-.025em b}\kern-.08em
    T\kern-.1667em\lower.7ex\hbox{E}\kern-.125emX}}
\begin{document}

\title{UGC-VIDEO: perceptual quality assessment of user-generated videos\\
}

\author{\IEEEauthorblockN{1\textsuperscript{st} Yang Li}
\IEEEauthorblockA{\textit{Institute of Digital Media, Peking University} \\
\textit{City University of Hong Kong}\\
liyang.00@pku.edu.cn}
\and
\IEEEauthorblockN{2\textsuperscript{nd} Shengbin Meng}
\IEEEauthorblockA{\textit{Bytedance Inc} \\
mengshengbin@bytedance.com}
\and
\IEEEauthorblockN{3\textsuperscript{rd} Xinfeng Zhang}
\IEEEauthorblockA{\textit{Department of Computer Science} \\
\textit{City University of Hong Kong}\\
xzhan44@cityu.edu.hk}
\and
\IEEEauthorblockN{4\textsuperscript{th} Shiqi Wang}
\IEEEauthorblockA{\textit{Department of Computer Science} \\
\textit{City University of Hong Kong}\\
shiqwang@cityu.edu.hk}
\and
\IEEEauthorblockN{5\textsuperscript{th} Yue Wang}
\IEEEauthorblockA{\textit{Bytedance (HK) Limited} \\
wangyue.v@bytedance.com}
\and
\IEEEauthorblockN{6\textsuperscript{th} Siwei Ma}
\IEEEauthorblockA{\textit{Institute of Digital Media} \\
\textit{Peking University}\\
swma@pku.edu.cn}
}

\maketitle

\begin{abstract}
Recent years have witnessed an ever-expanding volume of user-generated content (UGC) videos available on the Internet.
Nevertheless, progress on perceptual quality assessment of UGC videos still remains quite limited.
There are many distinguished characteristics of UGC videos in the complete video production and delivery chain, and one important property closely relevant to video quality is that there does not exist the pristine source after they are uploaded to the hosting platform, such that they often undergo multiple compression stages before ultimately viewed.
To facilitate the UGC video quality assessment, we created a UGC video perceptual quality assessment database.
It contains 50 source videos collected from TikTok with diverse content, along with multiple distortion versions generated by the compression with different quantization levels and coding standards. 
Subjective quality assessment was conducted to evaluate the video quality. Furthermore, we benchmark the database using existing quality assessment algorithms, and potential room is observed to future improve the accuracy of UGC video quality measures.
\end{abstract}

\begin{IEEEkeywords}
Video quality assessment, user-generated video, video compression
\end{IEEEkeywords}

\section{Introduction}
\label{sec:intro}
In traditional video production chains, video content was historically created and provided by a limited number of media producers, such as licensed broadcasters and production companies.
With the development of technologies such as multimedia and mobile networks, recent years have witnessed the explosion of user-generated content (UGC) and related services. 
UGC videos created by end users have several unique characteristics.
First, the lack of professional video capture equipment and proper shooting skills make the quality of UGC videos perceptually worse, which has been widely criticized by end-users.
Second, the low barriers in video production make the UGC content extremely diverse, and special effects are often incorporated to enhance the user experience.
Third, when uploading the compressed UGC videos to the  sharing platform, they often undergo another round of compression depending on the requirements of the hosting platform. 
As such, the artifacts from multiple rounds of compression are induced, while pristine videos are absent in the hosting platform. 
Nevertheless, in the literature, the progress on UGC video quality assessment is quite limited, and the traditional VQA measures may not accommodate these distinct properties.

To comprehensively evaluate the state-of-the-art VQA measures and promote further development and comparative analysis, 
extensive subjective quality evaluations of the UGC videos are important, which serve as the prerequisite for the research of objective quality evaluations.
There are several publicly available video databases for quality assessment. For example, LIVE~\cite{seshadrinathan2010study,seshadrinathan2010subjective} and LIVE Mobile~\cite{moorthy2012video,moorthy2012subjective} collect pristine source videos, which subsequently undergo various types of distortions.
In MCL-JCV~\cite{wang2016mcl}, a compressed video quality assessment database is created based on the just noticeable difference (JND) model.
Apparently, these databases with high quality source videos may not align with the UGC application scenarios as mentioned above.
There are also some databases that are more realistic in the UGC application scenario. LIVE-Qualcomm Mobile In-Capture~\cite{ghadiyaram2017capture} contains videos with a variety of complex distortions during the acquisition process, and
KoNViD-1k~\cite{hosu2017konstanz} is a subjectively annotated VQA database consisting of 1,200 public-domain video sequences.
Moreover, in~\cite{athar2017quality}, the author has studied the quality assessment of images undergoing multiple distortion stages.
Though these databases are more relevant to UGC videos, existing databases do not suffice to simulate the UGC production chain from acquisition to the processing on the hosting platform.
Motivated by the above observations, in this work, we create a new UGC video database which reflects the realistic scenarios: UGC-VIDEO.
This database contains 50 source videos from TikTok with a variety of the video content.
Regarding the distorted videos, we used two different compression standards to simulate the compression in the hosting platforms: H.264/AVC~\cite{wiegand2003overview} and H.265/HEVC~\cite{sullivan2012overview}.
Subjective test was conducted to evaluate the visual quality, based on which the comprehensive evaluations on objective models were further performed.

% The rest of this paper is organized as follows. The data collection procedure for the UGC-VIDEO database is described in Section~\ref{sec:databasecreation}.
% Section~\ref{sec:subjectiveexperiments} details the subjective testing and data analysis. 
% The performance of popular quality assessment models on this database is provided in Section~\ref{sec:benchmark}.
% Finally, concluding remarks and future work are given in Section~\ref{sec:conclusion}.

\section{UGC Database}
\label{sec:databasecreation}

The source videos in the database are intended to represent typical videos that users captured by their mobile phones and uploaded to TikTok.

\subsection{Video collection}
\label{ssec:videocollection}

To cover diverse content representing typical UGC videos, 
we began with a large-scale UGC video database containing 10000 videos, which were randomly selected from the videos uploaded to TikTok.
These videos are diverse in terms of shooting equipment, content, quality, resolution, duration, frame rate, etc.
Subsequently, from the large scale database, we selected videos according to the following principles:

\begin{itemize}
\item Duration longer than 10 seconds
\item Had a resolution with 720$\times$1280 (W$\times$H)
\item Played at around 30 frames per second (FPS)
% \item The scene belongs to one of the four selected categories (selfie, indoor, outdoor, screen content)
\end{itemize}

The selected videos share the same spatial resolution, and 720p is one of the most dominant UGC video formats on mobile phones.
Considering the diversity of video content, we further classified the videos into four categories, including selfie, indoor, outdoor and screen content.
For selfie videos, most of the areas are occupied by one or more people, and some of them have special effects.
Indoor and outdoor are common scenes, and indoor videos are usually shot close-up while outdoor videos are acquired from a distant view.
Screen content videos include recorded game video, recorded animation, etc. 
The filtered subset was further randomly sampled, leading to around 100 videos per category.

\subsection{Statistical study}
\label{ssec:attributecomputation}

To perform sampling from these 400 videos in a scientifically sound way, we study the content of the videos statistically based on three attributes, including spatial, temporal and blur.

As suggested in \cite{rec2008p}, spatial and temporal information of the scenes are critical factors for the level of impairment that is suffered when the scene is lossy transmitted. 

\emph{Spatial information:} The spatial information (SI) indicates the amount of spatial detail of a frame. Each frame $F_n$ is filtered by a Sobel filter, 
% then the standard deviation over the Sobel-filtered frame is computed. Finally,
the maximum standard deviation over all Sobel-filtered frames is regarded as the SI of the video, which is formulated in Eqn.~\eqref{eq:si}.
\begin{equation}
\label{eq:si}
SI=\max_{time}{\{std_{space}[Sobel(F_n)]}\}
\end{equation}

\emph{Temporal information:} The temporal information (TI) indicates the amount of temporal changes of a video. TI is derived based upon the motion difference $M_n(i,j)$, which represents the difference of pixel values between successive frames. As such, the maximum standard deviations of motion difference denotes the TI of the video, which is given by Eqn.~\eqref{eq:ti}.
\begin{equation}
\label{eq:ti}
TI=\max_{time}{\{std_{space}[M_n(i,j)]}\}
\end{equation}
\emph{Blur:} The acquired UGC videos are usually accompanied with varying degrees of blurring artifacts, which also significantly affects the video quality. Herein, the blur is assessed by the cumulative probability of blur detection (CPBD) metric~\cite{narvekar2011no}. We evenly extract one frame per second from the video, and the average CPBD value over these frames is regarded as the blur measurement.

\subsection{Video sampling}
\label{ssec:videosampling}

\begin{figure}[htbp]
\centering
\subfigure[]{
	\begin{minipage}[t]{0.46\linewidth}
	\centering
	\includegraphics[width=1\textwidth]{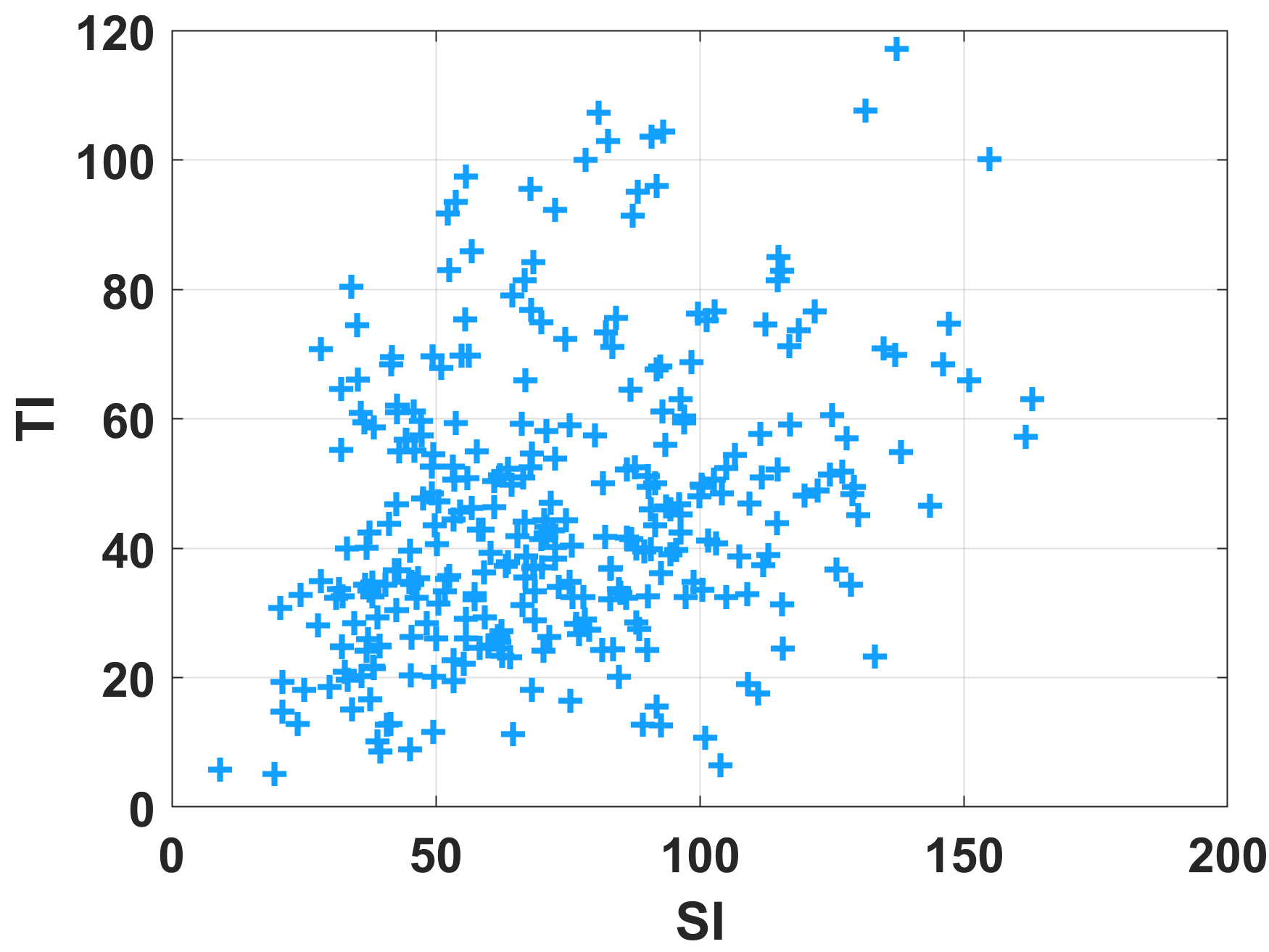}
	\label{Fig:siti400}
	\end{minipage}
}
\subfigure[]{
	\begin{minipage}[t]{0.46\linewidth}
	\centering
	\includegraphics[width=1\textwidth]{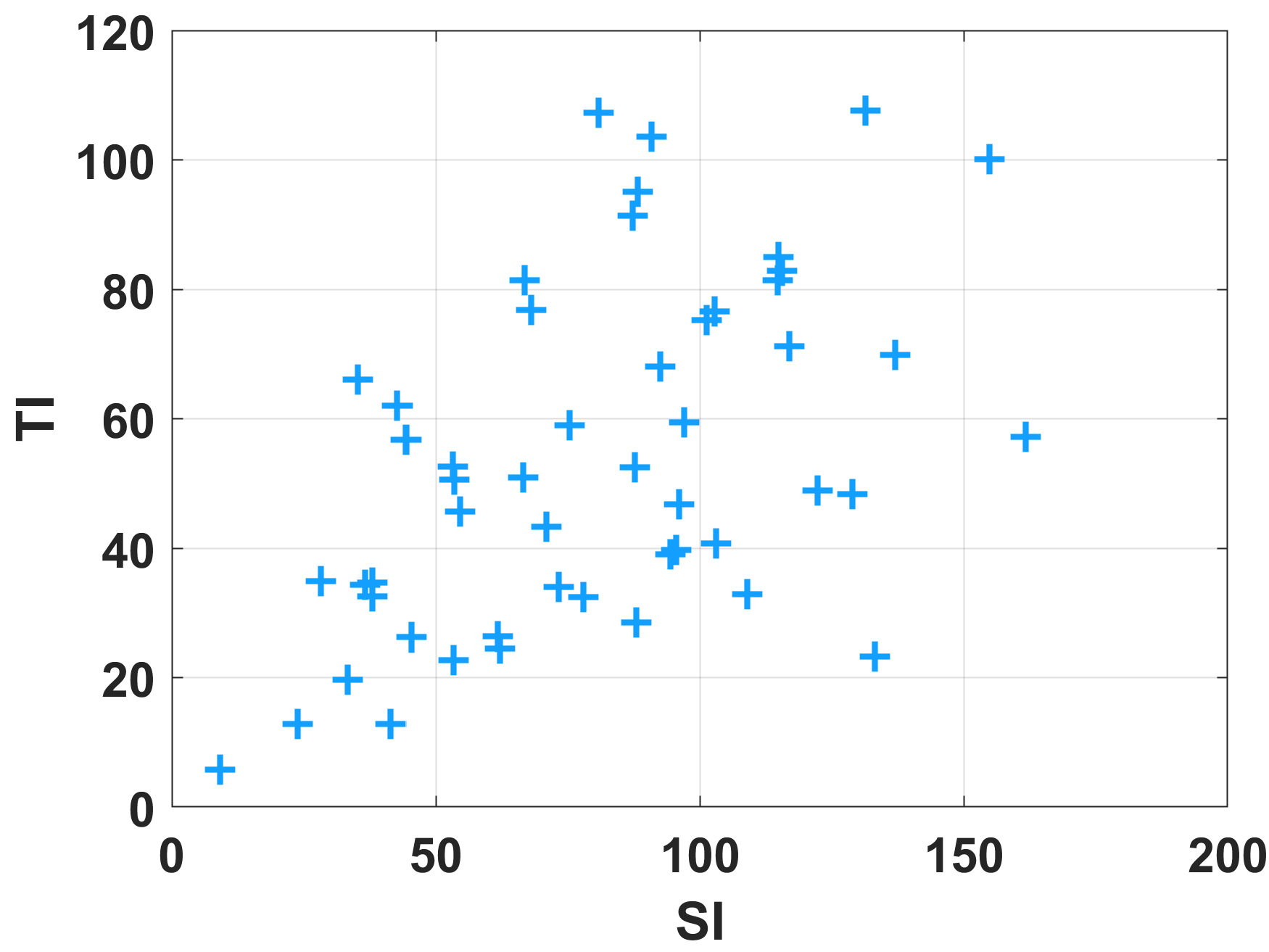}
	\label{Fig:siti50}
	\end{minipage}
}
\caption{Plot of SI and TI indices for videos. \subref{Fig:siti400} 400 videos before sampling. \subref{Fig:siti50} 50 videos after sampling.}
\label{Fig:SITI}
\end{figure}

From Fig.~\ref{Fig:siti400} we can see that these 400 videos are scattered in most areas of the feature space spanned by SI and TI, 
and the most densely distributed area is the area where both SI and TI are low.
To make the selected video uniformly distributed on calculated SI, TI and blur features, we use the sampling strategy introduced in ~\cite{vonikakis2016shaping}.
In particular let $S=\left\{ \bm{q}_i|\bm{q}_i\subset \mathbb{R},\bm{q}_i\sim D_S^M \right\}_{i=1}^K$ be the original database, while $M$ and $K$ denote the number of features and videos, respectively.
As such, we aim to select a subset $s=\left\{ \bm{\hat q}_i|\bm{\hat q}_i\subset \mathbb{R},\bm{\hat q}_i\sim D_s^M \right\}_{i=1}^N$ with the uniform distribution $\bm D\subset \mathbb{R}^{H\times M}$, 
each of its columns $\bm D_{*j}$ denoting the Probability Mass Function (PMF) across the $j^{th}$ dimension which is quantized into \(H\) bins.
Let $B=\left\{\bm{B}^m \right\}_{m=1}^M$ be a set of $M$ binary metrics, $b_{ij}^m$ denotes whether or not the $j^{th}$ item of $S$ belongs to $i^{th}$ interval of the target PMF for the dimension \(m\). And we introduce binary vector $\bm x\subset \mathbb{Z}_2^K$, where $x_i$ is decision variable determining whether $i^{th}$ item of $S$ belongs to subset $s$,
This problem can be formulated in Eqn.~\eqref{eq:opti}.
\begin{equation}
\label{eq:opti}
	\mathop{\min}_{\bm{x}} \sum_{m=1}^M\| \bm{B}^m \bm x - N\bm D_{*m}\|_1 \ \ s.t.\ \|\bm x\|=N
\end{equation}

\begin{figure}[t]
	\centering
	\includegraphics[width=0.85\columnwidth]{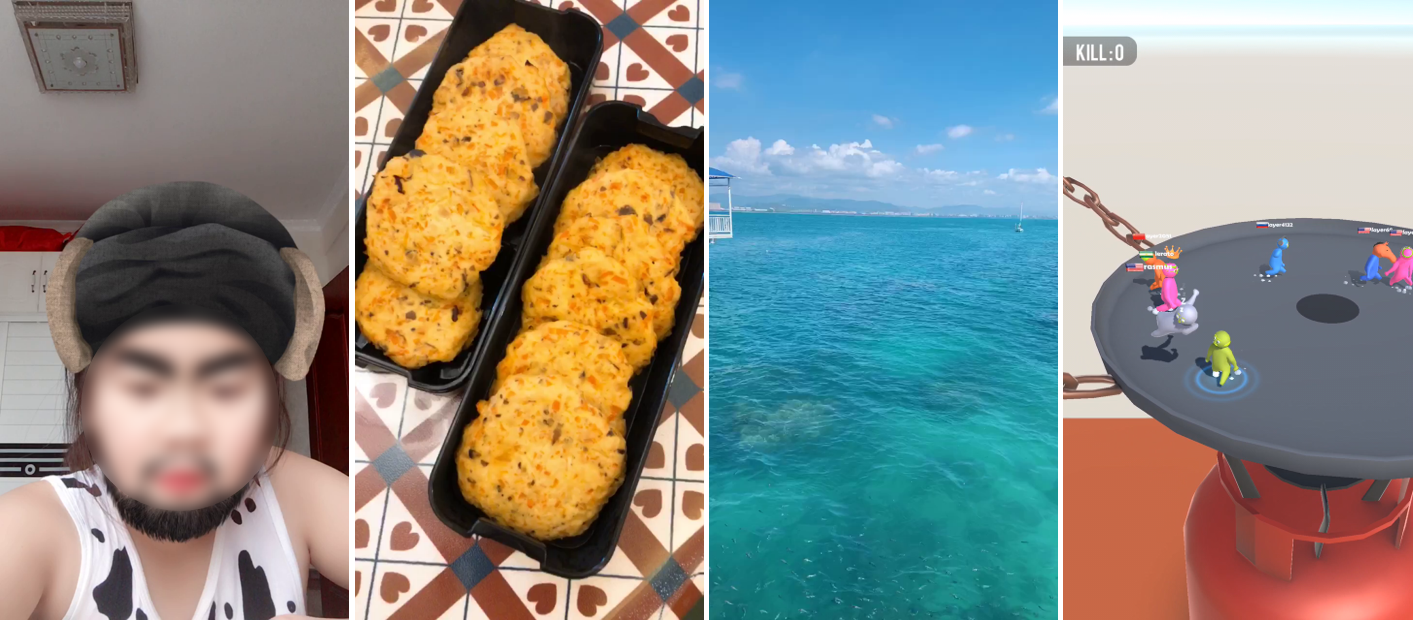}\\
	\caption{Selected source video sequences. The four images from left to right
	are extracted from selfie, indoor, outdoor and screen content video, respectively. The face has been blurred to show.}
	\label{Fig:examplevideo}
\end{figure}

By finding the optimal solution we can sample a subset from the original database that is nearly uniformly distributed over each feature.
We performed this operation in each video content category with $M=3$, $H=5$ and finally selected 12, 13, 13, 12 videos from selfie, indoor, outdoor and screen content videos, respectively.
And we extracted the first 10 seconds from these videos and removed the audio parts.
These selected 50 videos are well scattered in the feature space of SI and TI as shown in Fig.~\ref{Fig:siti50}. The snapshots of some selected videos are shown in Fig.~\ref{Fig:examplevideo}.

\subsection{Encoding configuration}
\label{ssec:distortedvideo}

Considering the fact that our primary goal of investigating the quality assessment of UGC videos for improving the video coding/transcoding performance. 
For each source, 10 corresponding sequences were generated by further encoding the video using two different codecs and five QPs.
Both H.264/AVC encoder x264~\cite{merritt2006x264} and H.265/HEVC encoder x265~\cite{x265} were used to encode each of the 50 sequences. The quality of the coded videos were determined by QPs with the values of 22, 27, 32, 37, 42 for both codecs. As such, with the inclusion of source videos, we have 50$\times$11 = 550 video sequences in the UGC-VIDEO database.

\section{SUBJECTIVE Evaluations}
\label{sec:subjectiveexperiments}

\subsection{Subjective testing}
\label{ssec:subtesting}

After collecting the videos, subjective testing was further conducted to obtain the subjective scores.
Single-stimulus (SS)~\cite{itu2012methodology} method, more specifically, a absolute category rating with hidden reference (ACR-HR)~\cite{itu1999subjective} paradigm was adopted to obtain opinion scores for the video database.
The subjects indicated the quality of each video he/she watched according the five-grade quality scales$-$``Excellent'',``Good'', ``Fair'', ``Poor'' and ``Bad'', corresponding to 5 to 1 points.

All the videos in our study were viewed by each subject.
To minimize the effects of viewer fatigue, we conducted the study in three sessions, each session lasted about half an hour, containing 16 or 17 source videos as well as all their corresponding compressed versions.
All test videos in a session were played one by one in a random order.
Besides, 10 ``dummpy presentations'' of various levels of distortions were introduced at the beginning of each session to stabilize the opinion of subjects, and the data issued from these presentations were not taken into account in the final results of the test.

A program was developed for this study on a Windows PC.
The videos were displayed one by one in their native resolution without scaling, and the subject needed to provide a video score by clicking the corresponding button within a few seconds after the video was played.
A Cathode Ray Tube (CRT) monitor with a display resolution of 1920$\times$1440 was used for display and the entire study was conducted using the same monitor. 
In total, 30 subjects participated in our study.

\subsection{Data analysis}
\label{ssec:dataanalysis}

We followed the procedure for screening of subjects as specified in ITU-R BT 500.13~\cite{itu2012methodology}.
% The kurtosis of scores are calculated to determine if the scores for each test presentation are normally distributed. Score range of each video is then computed as 2 or $\sqrt{20}$ standard deviations from the mean scores according to whether the scores are normally distributed. For each subject $i$, we count the number of scores above and blow this range, denoted as $P_i$ and $Q_i$.
% As such, the subject $i$ will be rejected if:
% \begin{equation}
% 	\frac {P_i+Q_i}{JK}>0.05 \ \ \text{and} \ \ |\frac{P_i-Q_i}{P_i+Q_i}|<0.3
% \end{equation}
% while $J$ and $K$ denote the number of distortion versions and source videos, respectively, and the corresponding values are 11 and 50.
No subjects were rejected at this stage in our study.
% Since this was a hidden-reference study, the reference videos of various quality were included in the set of the videos that the subject saw. 
Besides the mean opinion score (MOS), the differential mean opinion score (DMOS) can also be computed as the difference between the score that the subject provides for the source and the corresponding distorted videos. 

In essence, full reference quality evaluation algorithms can be greatly influenced by the quality of the reference video.
Fig.~\ref{Fig:psnrcompare} shows two sets of videos where the reference videos are of different quality. Moreover, these two distorted videos are compressed by H.265/HEVC, and their objective quality evaluations reflected solely from full reference measures such as PSNR and SSIM are quite close. However, their visual quality are significantly different, as noticeable blur can be seen in the compressed video from Fig.~\ref{Fig:psnr_compare1} while little noticeable distortion can be found in Fig.~\ref{Fig:psnr_compare2}.

\begin{figure}[t]
	\centering
	\includegraphics[width=0.8\columnwidth]{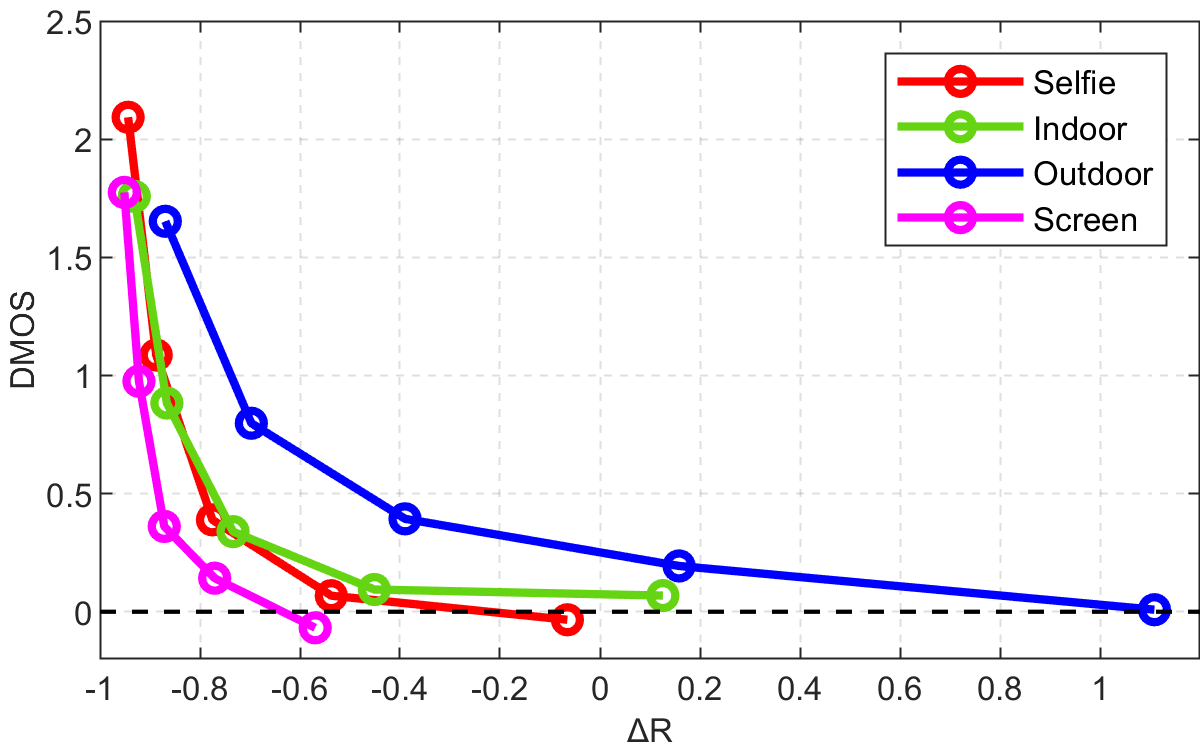}\\
	\caption{The relationship between DMOS and $\Delta{R}$ for each category.}
	\label{Fig:bit_dmos}
\end{figure}

The curves of DMOS as a function of $\Delta{R}$ in each category when transcoding by HEVC are plotted in Fig.~\ref{Fig:bit_dmos}, where
\begin{equation}
\label{eq:delta_bits}
\Delta{R} = \frac{R_{compressed} - R_{reference}}{R_{reference}}
\end{equation}
for each compressed video, $R_{reference}$ and $R_{compressed}$ denote the bit rate of reference video and compressed video, respectively. 
The $\Delta{R}$ and DMOS values of each point in the curve are the mean values of all videos under the same category and compression QP. Black dotted line indicates $DMOS=0$, the points below this line indicate that the quality of videos has been slightly improved after compression.
This is due to the fact that for some source videos with obvious distortion or noise, compression plays a role as smoothing or denoising.
As the QP increases, the most significant change in terms of $\Delta{R}-DMOS$ appears in outdoor videos,
while screen content videos are less affected.

\section{Evaluations of Objective Models}
\label{sec:benchmark}

\subsection{VQA for low quality reference videos}
\label{ssec:vqareference}
We evaluated the performance of several publicly available objective quality assessment algorithms based on subjective scores from the established database.
\begin{figure}[htbp]
\centering
\subfigure[]{
	\begin{minipage}[t]{0.9\linewidth}
	\centering
	\includegraphics[width=1\textwidth]{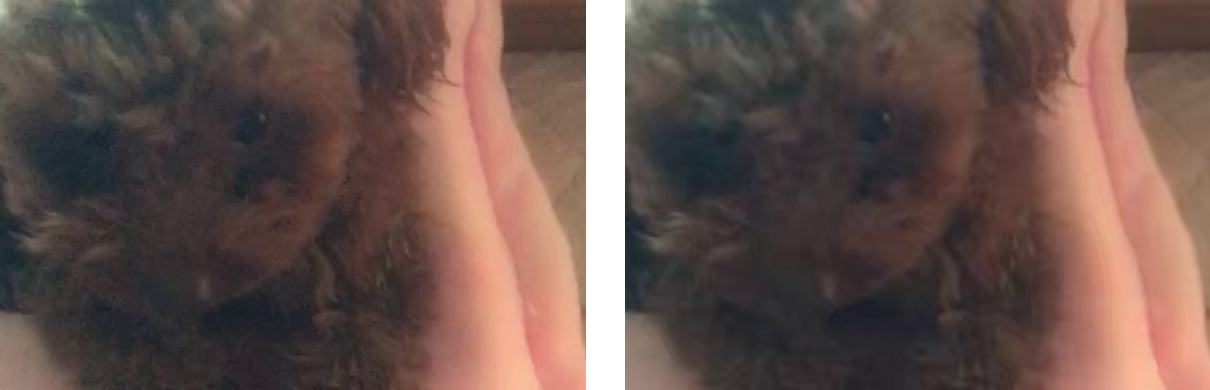}
	\label{Fig:psnr_compare1}
	\end{minipage}
}
\subfigure[]{
	\begin{minipage}[t]{0.9\linewidth}
	\centering
	\includegraphics[width=1\textwidth]{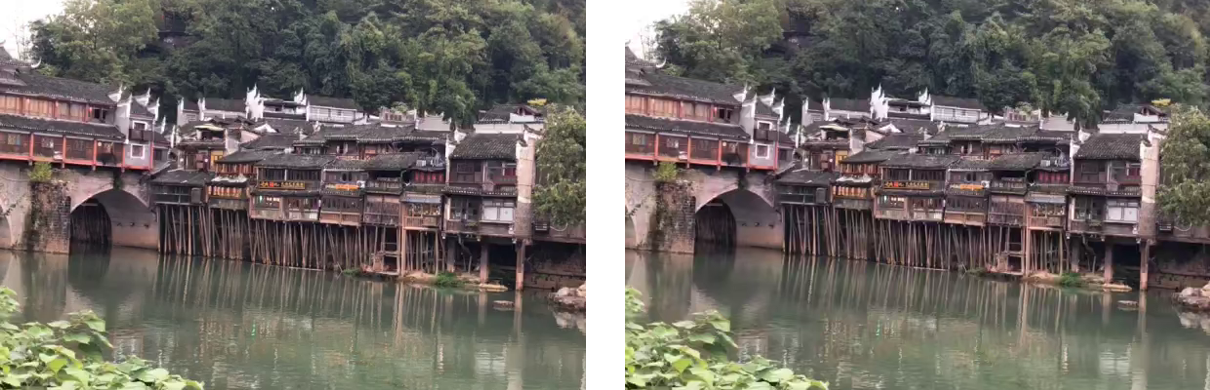}
	\label{Fig:psnr_compare2}
	\end{minipage}
}
\caption{Illustration of two sets of videos with high PSNR and SSIM values. Left: source videos; right: compressed videos. \subref{Fig:psnr_compare1} Left: MOS 3.21; Right: MOS 2.96, PSNR 41.73dB, SSIM 0.977. \subref{Fig:psnr_compare2} Left: MOS 4.43; right: MOS 4.11, PSNR 41.64dB, SSIM 0.996.}
\label{Fig:psnrcompare}
\end{figure}
For image quality assessment methods, they were extended to videos by averaging frame-level quality scores.
These evaluated full-reference and reduced-reference metrics include PSNR, SSIM~\cite{wang2004image},  VIF~\cite{sheikh2004image}, Multi-Scale SSIM (MS-SSIM)~\cite{wang2003multiscale}, SpEED-QA~\cite{bampis2017speed}, ViS3~\cite{vu2014vis3}, Video Multi-method Assessment Fusion (VMAF)~\cite{aaron2015challenges} and its phone screen viewing model VMAF-phone.

Three performance criteria including Spearman's rank ordered correlation coefficient (SROCC), Pearson's linear correlation coefficient (PLCC) and the root-mean-squared error (RMSE) were used in the evaluation.
PLCC and RMSE are computed after non-linear regression using a logistic function described in~\cite{sheikh2006statistical}, which is defined as,
\begin{equation}
f(x) = \beta_1(\frac{1}{2}-\frac{1}{1+e^{\beta_2(x-\beta_3)}})+\beta_4x+\beta_5.
\end{equation}

To further investigate the influence of low quality reference on the VQA algorithms,
the source videos are divided according to their 20th percentile value of all their MOS scores.
Then 110 compressed videos with low quality source are classified into the category 1, and the remaining 390 compressed videos are divided into category 2.
Table~\ref{tab:dmosobj} tabulates the SROCC, PLCC and RMSE between the algorithm scores and DMOS for each category, as well as across all videos. 
In principle, low quality reference will lead to poor performance of the reference quality measures. The results of this study support our conjecture, and we can see a significant performance degradation on videos with worse quality reference (category 1) comparing to videos with good reference (category 2) for almost all tested algorithms.
Also, by combining with the analysis of Fig.~\ref{Fig:psnrcompare} above, we can find that when the quality of the reference video is poor, the reference VQA algorithms cannot achieve the desired performance in terms of the correlations with either DMOS or MOS. 
%are not recommended, either directly assess the quality of the distorted video or evaluate the quality of the distorted video relative to the reference video.

\begin{table*}[bp]
\caption{Comparison of the performance of different quality assessment algorithms on DMOS. \subref{tab:dmossrocc} Spearman's rank order correlation; \subref{tab:dmosplcc} Pearson's Linear correlation coefficient; \subref{tab:dmosrmse} root-mean-squared-error}
\label{tab:dmosobj}
\centering  
\subtable{
	\begin{minipage}[t]{0.31\textwidth}
    \resizebox{1\columnwidth}{!}{
	\begin{tabular}{|c|c|c|c|}
	\hline
	Algorithm & Category 1  & Category 2  & All data \\ \hline
	PSNR      & 0.6181  & 0.7502  & 0.7351   \\ \hline
	SSIM      & 0.5444  & 0.8308  & 0.7708   \\ \hline
	MS-SSIM   & 0.5854  & 0.8318  & 0.7809   \\ \hline
	VIF       & 0.5622  & 0.7827  & 0.6895   \\ \hline
	SpEED-QA  & 0.7539  & 0.8183  & 0.7941   \\ \hline
	ViS3      & 0.8029  & 0.8455  & 0.8443   \\ \hline
    VMAF      & 0.7957  & 0.8662  & 0.8415   \\ \hline
	VMAF-phone& 0.7781  & 0.8695  & 0.8424   \\ \hline
	\end{tabular}}
	\label{tab:dmossrocc}
	\end{minipage}
}
\subtable{
	\begin{minipage}[t]{0.31\textwidth}
    \resizebox{1\columnwidth}{!}{
	\begin{tabular}{|c|c|c|c|}
	\hline
	Algorithm & Category 1  & Category 2  & All data \\ \hline
	PSNR      & 0.5852  & 0.7018  & 0.6749   \\ \hline
	SSIM      & 0.6735  & 0.8412  & 0.7979   \\ \hline
	MS-SSIM   & 0.6607  & 0.8420  & 0.8023   \\ \hline
	VIF       & 0.6129  & 0.8142  & 0.7023   \\ \hline
	SpEED-QA  & 0.7437  & 0.7229  & 0.8034   \\ \hline
	ViS3      & 0.8339  & 0.8463  & 0.8462   \\ \hline
    VMAF      & 0.8079  & 0.8999  & 0.8726   \\ \hline
	VMAF-phone& 0.7495  & 0.8375  & 0.8137   \\ \hline
	\end{tabular}}
	\label{tab:dmosplcc}
	\end{minipage}
}
\subtable{
	\begin{minipage}[t]{0.31\textwidth}
    \resizebox{1\columnwidth}{!}{
	\begin{tabular}{|c|c|c|c|}
	\hline
	Algorithm & Category 1  & Category 2  & All data \\ \hline
	PSNR      & 0.5421  & 0.5933  & 0.5998   \\ \hline
	SSIM      & 0.4943  & 0.4503  & 0.4899   \\ \hline
	MS-SSIM   & 0.5019  & 0.4492  & 0.4852   \\ \hline
	VIF       & 0.5283  & 0.4835  & 0.5787   \\ \hline
	SpEED-QA  & 0.4470  & 0.5754  & 0.4840   \\ \hline
	ViS3      & 0.3690  & 0.4436  & 0.4331   \\ \hline
    VMAF      & 0.3940  & 0.3632  & 0.3971   \\ \hline
	VMAF-phone& 0.4426  & 0.4550  & 0.4724   \\ \hline
	\end{tabular}}
	\label{tab:dmosrmse}
	\end{minipage}
}
\end{table*}

\subsection{VQA for videos under different content categories}
\label{ssec:vqacontent}

In the actual UGC transfer process, there is no straightforward and effective way to compute the quality of source videos. As suah, we can only regard these source videos with various quality as ``pristin'' references.
Herein, we evaluated the performance of algorithms for each video content category, and their correlation with MOS scores were computed.
Besides the above mentioned reference VQA algorithms,
no-reference assessment BRISQUE~\cite{mittal2011referenceless,mittal2012no}, NIQE~\cite{mittal2013making}, VIIDEO~\cite{mittal2015completely} and BLIINDS~\cite{saad2014blind} were also evaluated.
The SROCC, PLCC and RMSE results of four separate categories and the whole database are shown in Table~\ref{tab:benchmark}.
 
We can find that the existing algorithms may not provide reasonably accurate predictions on the UGC videos.
For most algorithms, they perform the worst on screen content videos, and the reason may be attributed to the fact that most algorithms do not consider the characteristics of this particular video content.
% For conventional measures PSNR and SSIM, there are obvious limitations when the source videos are diverse and already distorted.
No reference models BRISQUE and NIQE trained on natural images may not be appropriate for the the unnatural and artificially generated content.
Blind quality algorithms VIIDEO and BLIINDS also perform worse on all categories compared to reference algorithms.
These indicate that it is a great challenge to evaluate such diverse videos with various multi-stage distortions without any reference.
% The results exhibit that there is still large room for further improvement of UGC video quality assessment.
Besides, by comparing the results on ``All data'' in Table.~\ref{tab:dmosobj} and Table.~\ref{tab:benchmark}, 
We can find that most reference algorithms are more correlated with DMOS than with MOS.

\begin{table*}
\caption{Comparison of the performance of quality assessment algorithms on MOS. \subref{tab:srocc} Spearman's rank order correlation; \subref{tab:plcc} Pearson's Linear correlation coefficient; \subref{tab:rmse} root-mean-squared error}
\label{tab:benchmark}
\centering
\subtable{
	\begin{minipage}[t]{0.31\textwidth}
	\centering
    \resizebox{1\columnwidth}{!}{
	\begin{tabular}{|c|c|c|c|c|c|}
	\hline
	Algorithm & Selfie & Indoor & Ourdoor & Screen & All data \\ \hline
	PSNR      & 0.7149 & 0.7002 & 0.6643  & 0.4890 & 0.6120   \\ \hline
	SSIM      & 0.8423 & 0.7984 & 0.8571  & 0.4640 & 0.7140   \\ \hline
	MS-SSIM   & 0.8209 & 0.7833 & 0.8423  & 0.5071 & 0.7221   \\ \hline
	VIF       & 0.8372 & 0.8025 & 0.8065  & 0.6288 & 0.7361   \\ \hline
	NIQE      & 0.5109 & 0.4795 & 0.4529  & 0.1280 & 0.3144   \\ \hline
    BRISQUE   & 0.4364 & 0.3271 & 0.5801  & 0.3456 & 0.3543   \\ \hline
	SpEED-QA  & 0.8393 & 0.7471 & 0.8384  & 0.7460 & 0.7857   \\ \hline
	ViS3      & 0.7619 & 0.7059 & 0.8226  & 0.6985 & 0.7460   \\ \hline
	VMAF      & 0.8229 & 0.8208 & 0.8559  & 0.8252 & 0.8141   \\ \hline
	VMAF-phone& 0.8650 & 0.7926 & 0.8713  & 0.7890 & 0.8232   \\ \hline
	VIIDEO    & 0.1127 & 0.3484 & 0.2178  & 0.0255 & 0.0854   \\ \hline
	BLIINDS   & 0.3815 & 0.3859 & 0.0505  & 0.4621 & 0.1749   \\ \hline
	\end{tabular}}
	\label{tab:srocc}
	\end{minipage}
}
\subtable{
	\begin{minipage}[t]{0.31\textwidth}
	\centering
    \resizebox{1\columnwidth}{!}{
	\begin{tabular}{|c|c|c|c|c|c|}
	\hline
	Algorithm & Selfie & Indoor & Ourdoor & Screen & All data \\ \hline
	PSNR      & 0.7165 & 0.7327 & 0.6389  & 0.4517 & 0.5785   \\ \hline
	SSIM      & 0.8664 & 0.8466 & 0.8571  & 0.5901 & 0.7683   \\ \hline
	MS-SSIM   & 0.8445 & 0.8411 & 0.8651  & 0.6257 & 0.7726   \\ \hline
	VIF       & 0.8620 & 0.8204 & 0.8500  & 0.6334 & 0.6263   \\ \hline
	NIQE      & 0.5094 & 0.5110 & 0.5196  & 0.0560 & 0.1760   \\ \hline
    BRISQUE   & 0.4155 & 0.3462 & 0.6109  & 0.3283 & 0.3145   \\ \hline
	SpEED-QA  & 0.7477 & 0.6707 & 0.7297  & 0.7240 & 0.6725   \\ \hline
	ViS3      & 0.7867 & 0.7437 & 0.8724  & 0.7541 & 0.7832   \\ \hline
	VMAF      & 0.8843 & 0.8859 & 0.9071  & 0.8302 & 0.8633   \\ \hline
	VMAF-phone& 0.8609 & 0.8372 & 0.8650  & 0.7574 & 0.8069   \\ \hline
	VIIDEO    & 0.2512 & 0.1776 & 0.3259  & 0.0321 & 0.1567   \\ \hline
	BLIINDS   & 0.4147 & 0.4207 & 0.0004  & 0.4641 & 0.2160   \\ \hline
	\end{tabular}}
	\label{tab:plcc}
	\end{minipage}
}
\subtable{
	\begin{minipage}[t]{0.31\textwidth}
	\centering
    \resizebox{1\columnwidth}{!}{
	\begin{tabular}{|c|c|c|c|c|c|}
	\hline
	Algorithm & Selfie & Indoor & Ourdoor & Screen & All data \\ \hline
	PSNR      & 0.6091 & 0.5380 & 0.6162  & 0.7780 & 0.6836   \\ \hline
	SSIM      & 0.4361 & 0.4208 & 0.4126  & 0.7040 & 0.5364   \\ \hline
	MS-SSIM   & 0.4677 & 0.4276 & 0.4018  & 0.6803 & 0.5321   \\ \hline
	VIF       & 0.4426 & 0.4520 & 0.4220  & 0.6749 & 0.6533   \\ \hline
	NIQE      & 0.7514 & 0.6796 & 0.6843  & 0.8707 & 0.8250   \\ \hline
    BRISQUE   & 0.7943 & 0.7417 & 0.6341  & 0.8237 & 0.7956   \\ \hline
	SpEED-QA  & 0.5799 & 0.5864 & 0.5477  & 0.6015 & 0.6203   \\ \hline
	ViS3      & 0.5391 & 0.5285 & 0.3915  & 0.5728 & 0.5211   \\ \hline
	VMAF      & 0.4077 & 0.3668 & 0.3371  & 0.4862 & 0.4230   \\ \hline
	VMAF-phone& 0.4443 & 0.4324 & 0.4019  & 0.5694 & 0.4951   \\ \hline
	VIIDEO    & 0.8452 & 0.7780 & 0.7572  & 0.8716 & 0.8277   \\ \hline
	BLIINDS   & 0.7946 & 0.7172 & 0.8010  & 0.7725 & 0.8183   \\ \hline
	\end{tabular}}
	\label{tab:rmse}
	\end{minipage}
}
\end{table*}

\section{CONCLUSIONS}
\label{sec:conclusion}

We have introduced a UGC video database containing diverse UGC videos with subjective ratings.
The distinct properties of the database are that the sources videos are selected in a scientifically sound way, and the database was realistic in terms of the distortion process.
% The design principles and subjective testing procedures were detailed.
Full-reference, reduced-reference and no-reference quality assessment algorithms were validated on the proposed database.
The low correlations between subjective ratings and objective measures suggest that there is still large room to improve the quality assessment of UGC videos.
% The database is expected to advance the research of quality assessment and compression of UGC videos.
% In the future work, we will extend this database from multiple perspectives (e.g., content, compression methods, resolution).

\bibliographystyle{IEEEbib}
\bibliography{refs}

\end{document}